\documentclass[floatfix,showpacs,pra,aps,twocolumn]{revtex4}

 \usepackage{bm}
 \usepackage{amsmath}
 \usepackage{amssymb}
 \usepackage{latexsym}
 \usepackage{amsfonts}
 \usepackage{epsfig}
 \usepackage{psfrag}

 \newcommand{\ket}[1]{\left|#1\right>}
 \newcommand{\bra}[1]{\left<#1\right|}
 \newcommand{\braket}[2]
 {\left<#1|#2\right>}
 \newcommand{\nn}{\nonumber\\}
 
 \newcommand{\f}[1]{\mbox{\boldmath$#1$}}

 \newcommand{\bea}{\begin{eqnarray}}
 \newcommand{\ea}{\end{eqnarray}}
 \newcommand{\eea}{\end{eqnarray}}
 \newcommand{\ord}{{\cal O}}
 \newcommand{\abs}[1]{{\left| #1 \right|}}
 \newcommand{\HI}{H_{\rm I}}
 \newcommand{\HF}{H_{\rm F}}

\begin{document}

\title{Adiabatic preparation without Quantum Phase Transitions}

\author{Gernot Schaller}

\affiliation{Institut f\"ur Theoretische Physik, Hardenbergstra{\ss}e 36,
Technische Universit\"at Berlin, D-10623 Berlin, Germany}

\begin{abstract}
Many physically interesting models show a quantum phase transition when a single parameter 
is varied through a critical point, where the ground state and the first excited state become degenerate.
When this parameter appears as a coupling constant, these models can be understood as
straight-line interpolations between different Hamiltonians $\HI$ and $\HF$.
For finite-size realizations however, there will usually be a finite energy gap between ground and first excited state.
By slowly changing the coupling constant through the point with the minimum energy gap one thereby has 
an adiabatic algorithm that prepares the ground state of $\HF$ from the ground state of $\HI$.
The adiabatic theorem implies that in order to obtain a good preparation fidelity
the runtime $\tau$ should scale with the inverse energy gap and thereby also with the system size.
In addition, for open quantum systems not only non-adiabatic but also thermal excitations are likely to occur.
It is shown that -- using only local Hamiltonians -- for the 1d quantum Ising model and the cluster model 
in a transverse field the conventional straight line path can be replaced by a series of straight-line 
interpolations, along which the fundamental energy gap is always greater than a constant independent on the system size.
The results are of interest for adiabatic quantum computation since strong similarities between adiabatic quantum 
algorithms and quantum phase transitions exist. 
\end{abstract}

\pacs{
64.70.Tg, 
03.67.Bg, 
03.67.-a, 
03.67.Lx 
}

\maketitle


\section{Introduction}

The promises of quantum computers in solving certain difficult problems such as number factoring \cite{shor1997} and
database search \cite{grover1997a} have initiated a lot of research \cite{nielsen2000}.
As no quantum system can be isolated perfectly from the environment \cite{breuer2002}, 
the influence of decoherence is of great interest.
In the conventional scheme of quantum computation, such errors have to be taken into account
using expensive error correction schemes or by finding extremely fast quantum gates \cite{nielsen2000}.
This has drawn attention towards alternative computation schemes such as measurement-based \cite{raussendorf2001a,raussendorf2003a}, 
holonomic \cite{pachos2001a} or adiabatic \cite{farhi2001} quantum computation that either circumvent error-prone parts of the computation 
process or even provide some intrinsic robustness against imperfect implementation or decoherence.

The basic idea of adiabatic quantum computation (AQC) is to encode the solution to a difficult problem into the 
ground state of a problem Hamiltonian.
A suitable quantum system is prepared in the easily accessible ground state of a different initial Hamiltonian, which is
then slowly deformed into the problem Hamiltonian.
For slow evolutions the system will remain close to the instantaneous ground state and will therefore finally encode
the solution with high fidelity.
The adiabatic theorem relates the maximum evolution speed with the spectral properties of the time-dependent Hamiltonian \cite{sarandy2004}.
Adiabatic quantum computation is polynomially equivalent to the conventional scheme of quantum computation \cite{aharonov2007} and it is also 
believed to be somewhat robust to decoherent interactions at low reservoir temperatures \cite{childs2001}:
Intuitively, the ground state cannot decay and excitations can only occur for high temperatures.
This is also backed by more formal calculations:
For a time-independent system Hamiltonian and a thermalized reservoir with inverse temperature $\beta$, 
the thermalized system density matrix with exactly the same inverse temperature can 
(in Born-Markov and secular approximation) be shown to be a stationary state of the master equation 
for the reduced system density matrix \cite{breuer2002}.
If for system Hamiltonians with a slow time dependence these approximations are also valid \cite{childs2001,schaller2008a}, this
would imply robustness of the adiabatic scheme as long as the reservoir temperature
is smaller than the fundamental energy gap of the system.

Adiabatic quantum algorithms bear strong similarities to the dynamics through a quantum phase transition \cite{latorre2004a,orus2004a,sachdev2000}.
They connect very different (macroscopically distinguishable) ground states and typically, 
the fundamental energy gap is well controlled in the beginning and the end of the adiabatic algorithm.
Somewhere in between however, the fundamental gap may become very small and for nontrivial problems it will 
scale inversely with the system size \cite{znidaric2005,znidaric2006a}.
Around the minimum energy gap the ground state changes drastically \cite{zanardi2006a} and 
in the continuum limit $n\to\infty$, this would correspond to a quantum phase transition where an adiabatic evolution is impossible.
However, as one will for AQC always be interested in a finite-size system, the degeneracy between ground state and 
first excited state is usually replaced by an avoided crossing.
Here, the scaling of the energy gap with the system size is of great interest, as it will in practice mean a huge difference
whether the gap scales inverse exponentially in $n$ or merely polynomially.

In this paper, two simple models will be studied that conventionally connect two Hamiltonians through a straight-line interpolation,
along which in the continuum limit a quantum phase transition is encountered.
It will be shown that with using a nonlinear path between the two Hamiltonians, the phase transition can be avoided.


\section{1d transverse Ising model}\label{Sising}

The one-dimensional quantum Ising model in a transverse field \cite{sachdev2000} can be envisaged as a straight-line interpolation  
\bea\label{Elinear}
H(s) = (1-s) \HI + s \HF
\eea
between the initial Hamiltonian $\HI$ (at $s=0$) and the final Hamiltonian (at $s=1$)
\bea\label{Ehamising}
\HI = -\sum_{i=1}^n \sigma^x_i\,,\qquad 
\HF = -\sum_{i=1}^n \sigma^z_i \sigma^z_{i+1}\,,
\eea
where $\sigma^{x/z}_i$ denote the Pauli spin matrices acting on the $i^{\rm th}$
qubit and periodic boundary conditions $\sigma^z_{n+1}=\sigma^z_1$ are assumed, which is however not essential for the
scaling of the fundamental energy gap.
Both Hamiltonians in (\ref{Ehamising}) obey a 180 degree rotational symmetry around all $\sigma^x_i$-axes (bit-flip), which transforms $\sigma^z_i$ to $-\sigma^z_i$.

The ground state of the final Hamiltonian is two-fold degenerate, with 
\mbox{$\ket{\Psi_0^{(1)}} = \ket{0\ldots0}$} and 
\mbox{$\ket{\Psi_0^{(2)}}=\ket{1\ldots1}$},
where 
\mbox{$\sigma^z \ket{0}=+\ket{0}$} and 
\mbox{$\sigma^z \ket{1}=-\ket{1}$}.
As the eigenvalues of $\HF$ are proportional to the number of kinks in the basis states 
\mbox{$\sigma^z_i \ket{z_1\dots z_n}=(-1)^{z_i}\ket{z_1\dots z_n}$}, 
the two ground states are well separated from the excited states.

The unique ground state of the initial Hamiltonian in (\ref{Ehamising}) is given by
\bea\label{Etotalsup}
\ket{ {\cal S} } = \frac{1}{\sqrt{2^n}} \sum_{z=0}^{2^n-1} \ket{z} = \ket{\to} \otimes \ldots \otimes \ket{\to}\,,
\eea
where $\sigma^x \ket{\to} = +\ket{\to}$.
The excited states (with at least one spin flipped) are again well separated from the ground state.

Due to $\left[\HF,\bigotimes\limits_{\ell=1}^n \sigma^x_\ell\right]=0$, the bit-flip-parity is a conserved quantity 
during the interpolation (\ref{Elinear}), and since (\ref{Etotalsup}) is even under bit flip, the final
system state will for adiabatic evolution \cite{dziarmaga2005a,cincio2007a,mostame2007a} choose at $s=1$ the even subspace (symmetry breaking) with the 
huge Schr\"odinger cat ground state
\mbox{$\ket{\Psi_0^{\rm even}}=1/\sqrt{2}\left[\ket{0\ldots0}+\ket{1\ldots1}\right]$}, 
if the system is initialized in (\ref{Etotalsup}) at $s=0$.

Finally, note that the overlap between initial and final ground state is exponentially small in the system
size $n$.


\subsection{Straight-Line Interpolation}\label{SSsline}

At any point during the interpolation $s \in [0,1]$, the Hamiltonian $H(s)$ can be diagonalized: 
The successive application of Jordan-Wigner, Fourier and Bogoliubov 
transformations \cite{jordan1928a,sachdev2000} map the spin-1/2-Hamiltonian to a fermionic one, 
which (in the subspace of even bit-flip parity corresponding to even numbers of quasi-particles only) reads
\bea
H(s) = \sum_k \epsilon_k \left(\gamma_k^\dagger \gamma_k - \frac{1}{2}\right)\,,
\eea
with the fermionic operators $\gamma_k$ and single quasi-particle energies
\bea\label{Esinglepart}
\epsilon_k = \sqrt{1-4 \cos^2(ka/2)s(1-s)}\,,
\eea
see figure \ref{Fcomparison_ising} left panel.
The wavenumber $ka$ covers the range 
\mbox{$ka=\frac{\pi}{n}\left(1+2\mathbb{Z}\right) \qquad : \qquad \abs{ka}<\pi$}.

From equation (\ref{Esinglepart}) one obtains a minimum gap of 
\mbox{$g_{\rm min}=2\sin\left(\frac{\pi}{2n}\right)=\ord\{1/n\}$} 
and indeed one also observes for constant speed interpolation $s=t/\tau$ a quadratic scaling of the
adiabatic runtime $\tau_{\rm ad}=\ord\{n^2\}$ \cite{dziarmaga2005a} with the number of qubits $n$.
At the critical point $s_{\rm crit}=1/2$, the Ising model undergoes a second order quantum phase transition 
\cite{sachdev2000}, which manifests itself in a divergence of the second derivative of the ground state energy.


\subsection{Nonlinear Interpolation Path}

The linear interpolation scheme in equation (\ref{Elinear}) is rather simple but
not necessary for the adiabatic theorem to hold \cite{farhi0208135,siu2007a}.
The nonlinear interpolation we will consider here will connect the Hamiltonians of (\ref{Ehamising}) by a series
of straight-line interpolations along the Hamiltonians
\bea\label{Ehamseries_ising}
H_0 &=& - \sum_{i=1}^n \sigma^x_i = \HI\,,\nn
H_1 &=& - \sigma^z_1\sigma^z_2 - \sum_{i=3}^n \sigma^x_i\,,\nn
&\vdots&\nn
H_k &=& - \sum_{i=1}^{k} \sigma^z_i \sigma^z_{i+1} - \sum_{i=k+2}^n \sigma^x_i\,,\nn
&\vdots&\nn
H_{n-1} &=& - \sum_{i=1}^{n-1} \sigma^z_i \sigma^z_{i+1}\,,\nn
H_n &=& - \sum_{i=1}^{n} \sigma^z_i \sigma^z_{i+1} = \HF\,,
\eea
such that formally the time-dependent Hamiltonian is given by
\bea\label{Estepwise}
H(t) &=& \sum_{k=0}^{n-1} \Theta\left[s_k(t)\right] \Theta\left[1-s_k(t)\right]\times\nn
&\times&\left\{\left[1-s_k(t)\right] H_k + s_k(t) H_{k+1}\right\}\,,
\eea
where $\Theta(x)$ denotes the Heavyside step function and $s_k(t) = \frac{t-k \Delta t}{\Delta t}$ encodes a constant speed interpolation with
$n \Delta t=\tau$ such that 
\mbox{$s_k[(k+1)\Delta t] = 1= 1- s_{k+1}[(k+1)\Delta t]$} and $H(\tau) = \HF$.

Given a spin chain with (constant) $\sigma^z_i \sigma^z_{i+1}$-interactions, one has to apply a correspondingly stronger external field to
achieve an effective decoupling of the spins.
Physically the above nonlinear scheme could be approximated by a strong transverse magnetic field that within the distance between two spins rises linearly
from zero to maximum and then travels at constant speed along the spin chain. 
Conversely, one might imagine to slowly pull the spin chain out of a region with a strong stationary transverse field.
Recent proposals for quantum simulators of the Ising model in a transverse field \cite{porras2004a,mostame0803} however, even provide 
a more direct control of the local spin-spin interactions, such that all regions of the phase diagram can in principle be reached.

Note that the interpolation path (\ref{Ehamseries_ising}) does not destroy the bitflip symmetry, since $\left[H_k, \bigotimes\limits_{\ell=1}^n \sigma^x_\ell\right]=0$.
It is easy to see that the ground state of a Hamiltonian $H_k$ in (\ref{Ehamseries_ising}) (within the subspace of 
even bit-flip parity) is for $1 \le k \le n-1$ given by 
\bea
\ket{\Psi_k^{\rm 0, even}} = \frac{1}{\sqrt{2}}\left[\f{1} + \sigma^x_1 \ldots \sigma^x_{k+1}\right]{\cal H}_{k+2}\ldots{\cal H}_n
\ket{0\ldots0}\,,
\eea
where ${\cal H}_k=\frac{1}{\sqrt{2}}\left(\sigma^x_k + \sigma^z_k\right)$ denotes the Hadamard gate on qubit $k$, 
such that the overlap between two successive ground states yields
$\braket{\Psi_k^{\rm 0, even}}{\Psi_{k+1}^{\rm 0, even}}=1/\sqrt{2}$ for $0 \le k \le n-2$. 
In the last interpolation step, the ground state is even invariant $\braket{\Psi_{n-1}^{\rm 0, even}}{\Psi_{n}^{\rm 0, even}}=1$.
Hence, in every single step only slight transformations of the ground state are performed and intuitively, one may
expect that adiabatic preparation along this modified path should be more efficient than in the conventional scheme.

\begin{figure*}[ht]
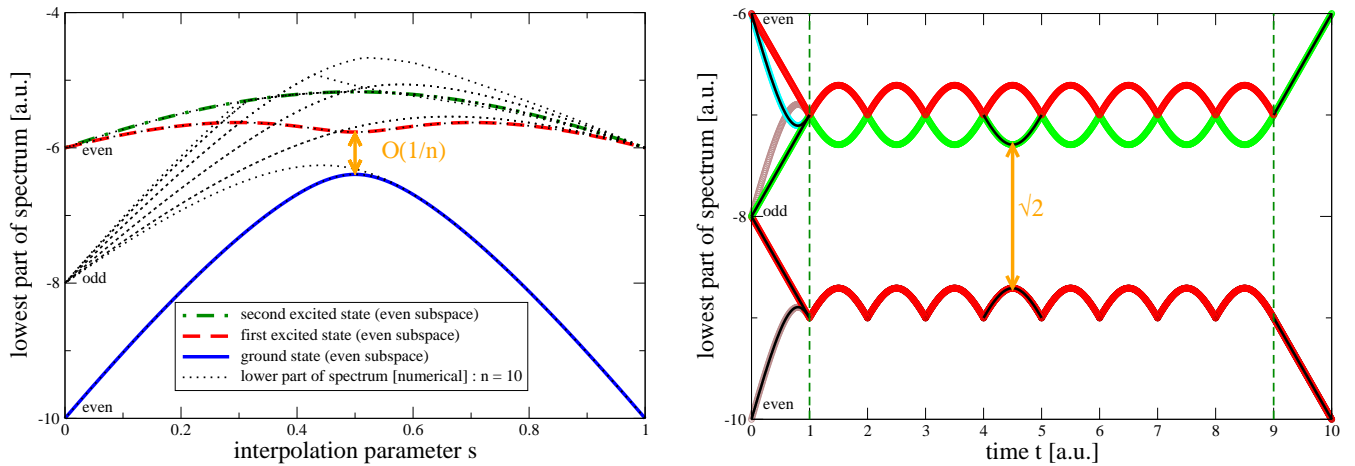

\includegraphics[height=6.1cm]{Ising_10.eps}\hspace{0.5cm}
\includegraphics[height=6.1cm]{Ising_stepwise_10.eps}
\caption{\label{Fcomparison_ising}[Color Online]
{\bf Left:}
Lowest eigenvalues for the 1d quantum Ising model in a transverse field (\ref{Ehamising}) and linear 
interpolation (\ref{Elinear}) for $n=10$. 
Thick lines show the analytical result for the eigenvalues of the even subspace (\ref{Esinglepart}), whereas the 
thin dotted lines give the numerical result \cite{arpack1998} for all lowest eigenvalues.
{\bf Right:}
Comparison of numerical calculation (dense symbols visible as thick lines) with the analytical prediction 
for $\kappa=0$ (solid black lines, only shown in intervals $[0,1]$, $[4,5]$, $[9,10]$)
for stepwise interpolation (\ref{Ehamseries_ising}) with $n=10$ and $\tau = 10$.
The different interpolation regimes are separated by dashed vertical lines.
Throughout the interpolation scheme the minimum gap in the even subspace is always larger than $\sqrt{2}$.
The ARPACK package \cite{arpack1998} only determines one part (e.g., the lower) of the spectrum, which is why not
all numerical eigenvalues are shown.
}
\end{figure*}

For the first interpolation step $H_0(s) \equiv (1-s)H_0 + s H_1$ in (\ref{Ehamseries_ising}), the first two qubits evolve 
independently from the rest of the system and it is straightforward to obtain the eigenvalues from the nontrivial contribution
of their four-dimensional subspace
\bea\label{Eevalfirst}
\lambda_0^\kappa &=& -\sqrt{5 s^2 - 8 s + 4} - (n-2) + 2\kappa \,,\nn
\lambda_1^\kappa &=& -s - (n-2) + 2\kappa\,,\nn
\lambda_2^\kappa &=& +s - (n-2) + 2\kappa\,,\nn
\lambda_3^\kappa &=& +\sqrt{5 s^2 - 8 s + 4} - (n-2) + 2\kappa\,,
\eea
where $\kappa \in \{0,1,\ldots,(n-2)\}$ counts the excitations resulting from qubits $3\ldots n$, 
see figure \ref{Fcomparison_ising} right panel.
By continuity we identify $\lambda_0^0$ and $\lambda_3^0$ as belonging to the even subspace and thus
one obtains in the first interpolation step for the minimum gap 
\mbox{$g_{\rm min}^{\rm even} = \min \left[\lambda_3^0-\lambda_0^0\right] = 4/\sqrt{5}$} at 
$s_{\rm min}=4/5$. 

For the intermediate steps we have
\bea\label{Estep}
H_k(s) &\equiv& (1-s)H_k + s H_{k+1}\nn
&=& - \sum_{i=1}^k \sigma^z_i \sigma^z_{i+1} - \sum_{i=k+3}^n \sigma^x_i\nn
&& - (1-s) \sigma^x_{k+2} - s \sigma^z_{k+1}\sigma^z_{k+2}
\eea
in (\ref{Ehamseries_ising}) for $1 \le k \le n-2$.
In order to obtain the spectrum conveniently, we will map the Hamiltonian using a suitable unitary transformation to another one
which can be diagonalized easily.
For the Ising model we consider the controlled NOT-gate
\bea\label{Ecnotgate}
S_{ij}^{\rm CNOT} = \frac{1}{2}\left(\f{1}_i + \sigma^z_i\right) \otimes \f{1}_j +\frac{1}{2}\left(\f{1}_i - \sigma^z_i\right) \otimes \sigma^x_j\,,
\eea
which is a Hermitian controlled-unitary operation \cite{nielsen2000}, where $i$ is the control and $j$ the target qubit.
It is straightforward to show that
\bea
S_{ij}^{\rm CNOT} \left[\sigma^z_i \otimes \f{1}_j \right] S_{ij}^{\rm CNOT} &=& \sigma^z_i \otimes \f{1}_j\,,\nn
S_{ij}^{\rm CNOT} \left[\sigma^z_i \otimes \sigma^z_j \right] S_{ij}^{\rm CNOT} &=& \f{1}_i \otimes \sigma^z_j\,,\nn
S_{ij}^{\rm CNOT} \left[\f{1}_i \otimes \sigma^x_j \right] S_{ij}^{\rm CNOT} &=& \f{1}_i \otimes \sigma^x_j\,.
\eea
Using the CNOT transformation at the transition region, the Hamiltonian (\ref{Estep}) is mapped to
\bea
S_{k+1, k+2}^{\rm CNOT} H_k(s) S_{k+1, k+2}^{\rm CNOT} &=&  - \sum_{i=1}^k \sigma^z_i \sigma^z_{i+1} - \sum_{i=k+3}^n \sigma^x_i\nn
&& - (1-s) \sigma^x_{k+2} - s \sigma^z_{k+2}\,,
\eea
where it is visible that the qubit sets 
\mbox{$\{1, \ldots, (k+1)\}$}, 
\mbox{$\{(k+2)\}$}, and 
\mbox{$\{(k+3), \ldots, n\}$} are mutually decoupled.
Evidently, one obtains for qubits 
\mbox{$1\ldots (k+1)$} just the eigenvalues of the Ising model with open boundary conditions 
(i.e., with minimum energy $-k$ for a two-fold degenerate ground state and fundamental energy gap 2)
and for qubits 
\mbox{$(k+3), \ldots, n$} a minimum energy of $-(n-k-2)$ for the unique ground state and a fundamental energy gap of 2.
The nontrivial part of the spectrum arises from the subspace of qubit $(k+2)$, where one obtains for the eigenvalues 
\bea\label{Eevalmed}
\lambda_{\pm}^\kappa = \pm \sqrt{1-2s(1-s)} -(n-2) + 2\kappa\,,
\eea
where $\kappa \in \{0,1,\ldots,(n-2)\}$, see figure \ref{Fcomparison_ising} right panel.
Therefore, the fundamental energy gap equates to $g_{\rm min}^{\rm even}= \sqrt{2}$, which is independent on the system size.

The last step is trivial: As $\left[H_{n-1}, H_n\right]=0$, both Hamiltonians can be diagonalized simultaneously and the spectra of 
$H_{n-1}$ and $H_n$ are therefore just connected by straight lines as is also visible in figure \ref{Fcomparison_ising} right panel.

Therefore, we obtain with the new interpolation path a lower bound on the minimum fundamental energy gap
\bea
g_{\rm min}^{\rm even} \ge \sqrt{2}
\eea
in the even subspace.
Note that as a sanity check, in figure \ref{Fcomparison_ising} right panel a perfect agreement is found between analytically and 
numerically calculated eigenvalues \cite{arpack1998} is found.

In the absence of a reservoir one would have an adiabatic scheme to efficiently prepare large Schr\"odinger cat states:
If the quantum system is initialized in its initial ground state
(\ref{Etotalsup}) (by applying a strong magnetic field) and then slowly evolved through the 
series of interpolations towards $\HF$, the adiabatic theorem will guarantee
that the system will remain close to its instantaneous ground state.
Evidently, the last step in (\ref{Ehamseries_ising}) can also be omitted without changing 
the ground state, just the energies will differ.
This does not mean that one can adiabatically deform $\HI$ in $\HF$ in (\ref{Ehamising}) in constant time, 
as the number of steps in (\ref{Estepwise}) scales linearly with the system size $n$.
In order to obtain a fixed final fidelity of the desired state 
$\ket{\Psi_{\rm final}}=1/\sqrt{2}\left[\ket{0\ldots0}+\ket{1\ldots1}\right]$,
the adiabatic runtime $\tau_{\rm ad}$ will scale nearly linearly with the system size $\tau_{\rm ad} \propto n \log n$ 
for the new scheme (\ref{Ehamseries_ising}) instead of quadratically \cite{dziarmaga2005a} for the straight-line interpolation scheme (\ref{Elinear}).
The logarithmic corrections are expected from the increasing number of degeneracies in the first excited states \cite{schaller2006a}.
However, this favorable scaling could also have been achieved by using the conventional scheme (\ref{Elinear}) with
straight-line but adaptive-speed interpolation \cite{schaller2006a,jansen2007a}.
Therefore, the advantage of the nonlinear scheme is rather given by the fact that constant-speed interpolation suffices 
and -- more importantly -- for couplings to a reservoir that do not destroy the bitflip symmetry the new scheme
provides some resistance to thermal excitations as long as $k_{\rm B} T_{\rm res} < \ord\{1\}$.

For realistic systems however it is to be expected that the conservation of bit-flip parity
will be destroyed by the existence of a reservoir, such that amplitude from the even subspace
may leak into the subspace of odd bitflip parity \cite{mostame2007a,fubini2007a}
(such effects may also limit the robustness of other models against decoherence, see e.g. \cite{balachandran2008a}).
The net effect would be the decay of the desired Schr\"odinger cat state.
In principle, this could be avoided with an additional (time-independent) Hamiltonian of the form
\bea
\Delta \HF &=& \frac{\alpha}{2}\left[\f{1} - \bigotimes\limits_{\ell=1}^n \sigma^x_\ell\right]\,,\nn
\eea
to safely separate the even and odd subspaces.
Unfortunately, it involves $n$-qubit interactions and is thus presumably difficult to implement.
In the following, we will therefore consider a model with a local Hamiltonian that has a unique ground state.


\section{Preparation of the Cluster State}\label{Scluster}

The cluster state on a lattice ${\cal L}$ can be encoded in the ground state of 
the Hamiltonian
\bea\label{Ehamcluster}
H = - \sum_{\mu \in {\cal L}} \sigma^x_\mu \bigotimes_{\nu \stackrel{\cal L}{\sim} \mu} \sigma^z_\nu\,,
\eea
where $\nu \stackrel{\cal L}{\sim} \mu$ denotes all sites $\nu$ that are connected to the site $\mu$ in the lattice $\cal L$ by a link.
In order to see that the cluster state is the ground state of (\ref{Ehamcluster}), we use the
controlled Z-gate
\bea\label{Ecsigmaz}
S_{ij}^{\rm CZ} = \frac{1}{2}\left(\f{1}_i + \sigma^z_i\right) \otimes \f{1}_j +\frac{1}{2}\left(\f{1}_i - \sigma^z_i\right) \otimes \sigma^z_j = S_{ji}^{\rm CZ}\,,
\eea
which acts as 
\bea
S_{ij}^{\rm CZ} \ket{z_i}\otimes \ket{z_j} = (-1)^{z_i z_j} \ket{z_i}\otimes \ket{z_j}
\eea
in the computational basis.
Obviously, it is Hermitian and $\left[S_{ij}^{\rm CZ}, S_{kl}^{\rm CZ}\right]=0$, such that we can unambiguously define the
global unitary operation
\bea\label{Esglobal}
S_{\cal L} = \prod_{\nu \stackrel{\cal L}{\sim} \mu} S_{\mu\nu}^{\rm CZ} = S_{\cal L}^\dagger\,.
\eea
With using the relations (see also \cite{doherty08024314})
\bea
S_{ij}^{\rm CZ} \left[\sigma^z_i \otimes \sigma^x_j \right] S_{ij}^{\rm CZ} &=& \f{1}_i \otimes \sigma^x_j\,,\nn
S_{ij}^{\rm CZ} \left[\sigma^z_i \otimes \f{1}_j \right] S_{ij}^{\rm CZ} &=& \sigma^z_i \otimes \f{1}_j\,,\nn
S_{ij}^{\rm CZ} \left[\f{1}_i \otimes \sigma^z_j \right] S_{ij}^{\rm CZ} &=& \f{1}_i \otimes \sigma^z_j
\eea
it is easy to see that the unitary transformation (\ref{Esglobal}) maps the Hamiltonian (\ref{Ehamcluster}) into
decoupled spins
\bea\label{Edecoupled}
S_{\cal L} H S_{\cal L}^\dagger = - \sum_{\mu \in {\cal L}} \sigma^x_\mu\,.
\eea
This implies that the ground state of (\ref{Ehamcluster}) is unique and separated by a safe energy gap 
of $\Delta E=2$ from the first excited states.
From the ground state of (\ref{Edecoupled}) it can be obtained via
\bea
\ket{\Psi_{\rm cl}} &=& S_{\cal L} \ket{\to}\otimes \ldots \otimes \ket{\to}\nn
&=& \frac{1}{\sqrt{2^n}} \sum_{z=0}^{2^n-1} (-1)^{\sum\limits_{\mu\nu} L_{\mu\nu} z_\mu z_\nu} \ket{z_1\ldots z_n}\,,
\eea
where $L_{\mu\nu}=1$ if $\mu \stackrel{\cal L}{\sim} \nu$ and $L_{\mu\nu}=0$ otherwise, and $n=\abs{\cal L}$ denotes the
number of lattice sites.
Above equation already gives the recipe of preparing the cluster state using only the two-qubit unitary operations (\ref{Ecsigmaz}).
However, considering the destructive influence of an environment it would be desirable to encode the
cluster state in the ground state of a Hamiltonian, which could then be reached by adiabatic transformation, see below.


\subsection{The 1d cluster state}

\begin{figure*}[ht]
\includegraphics[height=6.1cm]{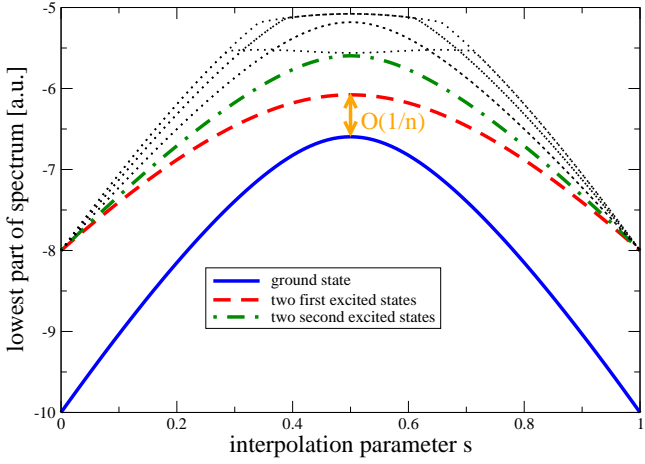}\hspace{0.5cm}
\includegraphics[height=6.1cm]{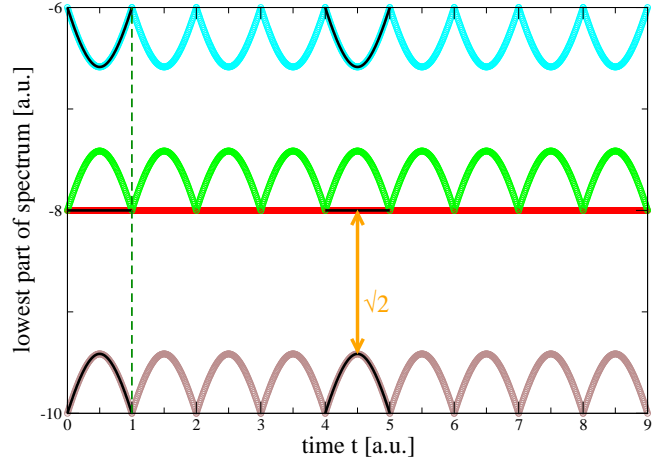}
\caption{\label{Fcomparison_cluster}[Color Online]
{\bf Left:}Numerical Calculation of the lowest eigenvalues for the adiabatic cluster model in a transverse field (\ref{Ehamcluster}) for $n=10$. 
Thick lines show the lowest five eigenvalues.
{\bf Right:}Numerical Calculation of the lowest eigenvalues for the stepwise cluster model (\ref{Ehamseries_cluster1d}) for $n=10$ and $\Delta t=1$. 
Analytical predictions (solid lines) resulting from (\ref{Eeval_cluster}) for $\kappa=0$ perfectly match the numerical results.
}
\end{figure*}

In analogy to the Ising model in a transverse field (\ref{Ehamising}), one could prepare the cluster 
state on a 1d spin chain adiabatically by a straight line interpolation (\ref{Elinear}) between
\bea\label{Ehamcluster1d}
\HI &=& -\sum_{i=1}^n \sigma^x_i\qquad \mbox{and}\nn
\HF &=& -\sigma^x_1 \sigma^z_2 - \left[\sum_{i=2}^{n-1} \sigma^z_{i-1} \sigma^x_i \sigma^z_{i+1}\right] - \sigma^z_{n-1} \sigma^x_n\,,
\eea
what is referred to as 'transverse field cluster model' in \cite{doherty08024314}.
It has been shown that the above model exhibits a quantum phase transition at $s_{\rm crit}=1/2$ \cite{pachos2004a},
which can be understood as (\ref{Ehamcluster1d}) can be mapped via a duality transformation to two Ising models in a 
transverse field \cite{doherty08024314}.
Accordingly, this kind of preparation would be vulnerable to nonadiabatic and thermal excitations in the large $n$ limit, as the 
minimum energy gap will scale as $\ord\{1/n\}$,
see also figure \ref{Fcomparison_cluster} left panel.

\begin{figure}[ht]
\includegraphics[height=2.5cm]{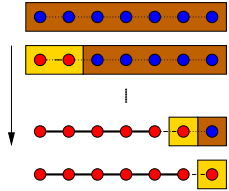}
\caption{\label{Flattice1d}[Color Online]
Stepwise buildup of the cluster state on a 1d spin chain with $n=7$ spins within 6 steps using a runtime of $\tau = \ord\{n\}$. 
In lattice ${\cal L}_1$, all spins are unconnected (dark-blue) and in lattice ${\cal L}_2$ only two spins are connected (grey-red) by
solid lines.
Thin dotted lines show the intended order of inclusion in the connected sublattice.
Dark shaded regions (brown) show the presence of a strong transverse field, whereas in the light-shaded regions (yellow) the field is gradually 
decreased from maximum to zero.
}
\end{figure}

Therefore, we will consider the stepwise interpolation scheme (\ref{Estepwise}) along 
\bea\label{Ehamseries_cluster1d}
H_0 &=& - \sum_{i=1}^n \sigma^x_i = \HI\,,\nn
H_1 &=& - \sigma^x_1\sigma^z_2 - \sigma^z_1 \sigma^x_2 - \sum_{i=3}^n \sigma^x_i\,,\nn
H_2 &=& - \sigma^x_1\sigma^z_2 -\sigma^z_1 \sigma^x_2 \sigma^z_3 - \sigma^z_2 \sigma^x_3 - \sum_{i=4}^n \sigma^x_i\,,\nn
&\vdots&\nn
H_k &=& - \sigma^x_1\sigma^z_2 - \sum_{i=2}^k \sigma^z_{i-1} \sigma^x_i \sigma^z_{i+1} - \sigma^z_k \sigma^x_{k+1} - \sum_{i=k+2}^n \sigma^x_i\nn
&=& -\sum_{\mu \in {\cal L}_k} \sigma^x_\mu \bigotimes_{\nu \stackrel{{\cal L}_k}{\sim} \mu} \sigma^z_\nu\,.
\eea
The physical implementation of such a model would of course be more demanding: 
Given that the necessary three-body interactions can be simulated by logical qubits (possibly built of several physical ones 
along the lines of \cite{bartlett2006a}), the region of a non-vanishing transverse field gradient should here also be confined 
within the distance between two logical qubits.
In order to simplify the notation, we have written the intermediate Hamiltonian of (\ref{Ehamseries_cluster1d}) as
in (\ref{Ehamcluster}) and just modify the number of existent links in the lattices ${\cal L}_k$, see figure \ref{Flattice1d}.
Then, we obtain for an intermediate interpolation step
\bea
H_k(s) &\equiv & (1-s) H_k + s H_{k+1}\nn
&=& - \sigma^x_1 \sigma^z_2 - \sum_{i=2}^{k} \sigma^z_{i-1} \sigma^x_i \sigma^z_{i+1} - \sum_{i=k+3}^n \sigma^x_i\nn
&& - (1-s) \sigma^z_k \sigma^x_{k+1} - (1-s) \sigma^x_{k+2}\nn
&& - s \sigma^z_{k} \sigma^x_{k+1} \sigma^z_{k+2} - s \sigma^z_{k+1} \sigma^x_{k+2}\,,
\eea
where it becomes visible that by applying the global unitary only on the existing links in lattice ${\cal L}_k$, the
model is again decoupled nearly completely
\bea
S_{{\cal L}_k}  H_k(s) S_{{\cal L}_k}^\dagger &=& -\sum_{i=1\; : \; i \notin \{(k+1),(k+2)\} }^n \sigma^x_i\nn
&&-s \left(\sigma^x_{k+1} \sigma^z_{k+2} + \sigma^z_{k+1} \sigma^x_{k+2}\right)\nn
&&-(1-s) \left(\sigma^x_{k+1} + \sigma^x_{k+2}\right)\,.
\eea
In above equation, only qubits $\{(k+1),(k+2)\}$ contribute the nontrivial parts of the spectrum and we obtain for the eigenvalues
\bea\label{Eeval_cluster}
\lambda_1^\kappa &=& -2\sqrt{1-2s(1-s)} - (n-2) + 2\kappa\,,\nn
\lambda_2^\kappa &=& \lambda_3^\kappa = -(n-2)+2\kappa\,,\nn
\lambda_4^\kappa &=& +2\sqrt{1-2s(1-s)}-(n-2)+2\kappa
\eea
with $\kappa \in \{0,1,\ldots,(n-2)\}$.
Note that the first interpolation step yields the same eigenvalues, as is also 
visible in the numerical calculation of the spectrum, see figure \ref{Fcomparison_cluster} right panel.


\subsection{The 2d cluster state}

In 2d, the cluster state can serve as a universal resource for measurement-based quantum computation \cite{raussendorf2001a,raussendorf2003a}.
Therefore, we will here also state some known facts about its adiabatic preparation via a straight-line 
interpolation (\ref{Elinear}) between the Hamiltonians
\bea
\HI = -\sum_{\mu \in {\cal L}} \sigma^x_\mu\,,\qquad 
\HF = -\sum_{\mu \in {\cal L}} \sigma^x_\mu \bigotimes_{\nu \stackrel{{\cal L}}{\sim} \mu} \sigma^z_\nu\,,
\eea
where ${\cal L}$ denotes a square lattice, such that five-body interactions are required in order 
to implement $\HF$.
For the direct straight-line interpolation one can map the above model with a duality transformation \cite{doherty08024314} 
to one with a known quantum phase transition at $s_{\rm crit}=1/2$ \cite{xu2004a}.

\begin{figure}[ht]
\includegraphics[height=2.5cm]{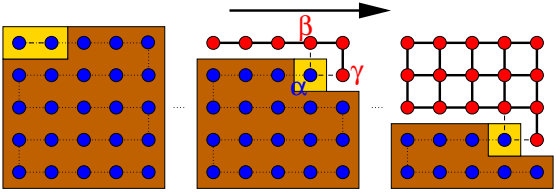}
\caption{\label{Flattice2d}[Color Online]
Stepwise buildup of the cluster state on a 2d spin lattice with $n=25$ spins in 24 steps.
Thin dotted lines show the intended order of inclusion in the connected sublattice.
In each single step, either only one or two links are added.
With regard to realization, one could imagine a region with a strong 
transverse magnetic field (dark-brown shaded region), which is turned off slowly (light-yellow shaded region)
along a rectangular zig-zag trajectory as outlined by the thin dotted lines.
}
\end{figure}

It is easy to see that one can choose an interpolation scheme where either only one or two links
are added at once, see figure \ref{Flattice2d}.
Whenever only a single link is inserted, we have the same situation as in the 1d case, 
therefore we will consider the more interesting case of adding two links at once here.
Suppose a node $\alpha$ is unconnected in lattice ${\cal L}_k$ (i.e., at $s=0$ with Hamiltonian $H_k$) and shall be
connected to two other nodes $\beta$ and $\gamma$ in lattice ${\cal L}_{k+1}$ (i.e., at $s=1$ with Hamiltonian $H_{k+1}$).
Suppose also that the nodes $\beta$ and $\gamma$ are already connected to other nodes in lattice ${\cal L}_k$ as depicted in figure \ref{Flattice2d}.
Then, the corresponding Hamiltonian can be written as
\bea
H_k(s) &\equiv & (1-s) H_k + s H_{k+1}\nn
&=& - \sum_{\mu \neq \alpha\beta\gamma} \sigma^x_\mu \bigotimes_{\nu \stackrel{{\cal L}_k}{\sim}\mu} \sigma^z_\nu\nn
&&-s\left(\sigma^x_\alpha \sigma^z_\beta \sigma^z_\gamma 
+ \sigma^z_\alpha \sigma^x_\beta \bigotimes_{\nu \stackrel{{\cal L}_k}{\sim} \beta} \sigma^z_\nu 
+ \sigma^z_\alpha \sigma^x_\gamma \bigotimes_{\nu \stackrel{{\cal L}_k}{\sim} \gamma} \sigma^z_\nu \right)\nn
&&-(1-s)\left(\sigma^x_\alpha + \sigma^x_\beta \bigotimes_{\nu \stackrel{{\cal L}_k}{\sim} \beta} \sigma^z_\nu + \sigma^x_\gamma \bigotimes_{\nu \stackrel{{\cal L}_k}{\sim} \gamma} \sigma^z_\nu \right).
\eea
Using the unitary transformation $S_{{\cal L}_k}$, we again arrive at a nearly decoupled model (where only three spins are mutually coupled)
\bea
S_{{\cal L}_k} H(s) S_{{\cal L}_k}^\dagger &=& - \sum_{\mu \neq \alpha\beta\gamma} \sigma^x_\mu 
- (1-s) \left(\sigma^x_\alpha + \sigma^x_\beta + \sigma^x_\gamma\right)\nn
&& - s \left(\sigma^x_\alpha \sigma^z_\beta \sigma^z_\gamma + \sigma^z_\alpha \sigma^x_\beta + \sigma^z_\alpha \sigma^x_\gamma\right)\,.
\eea
Evaluating the eigenvalues of the $8\times 8$ matrix generated by qubits \mbox{$\{\alpha,\beta,\gamma\}$} yields (for simplicity we only give the two lowest eigenvalues)
\bea
\lambda_0^\kappa &=& -\sqrt{5-10s(1-s)+2\sqrt{4(1-2s)^2+25 s^2 \left(1-s\right)^2}}\nn
&&-(n-3)+2\kappa\,,\nn
\lambda_1^\kappa &=& -1-(n-3)+2\kappa\,,
\eea
where $\kappa \in \{0,1,\ldots, (n-3)\}$.
Therefore, we obtain a fundamental energy gap of 
\mbox{$g_{\rm min}=\sqrt{5}-1$} at $s_{\rm min}=1/2$, which is independent on the system size.

Note that we have so far allowed the external field to vary at a single site on the lattice only.
If one does not have this constraint many of the steps can be performed in parallel, such that
the overall time required to generate the cluster state is further reduced, compare also \cite{siu2007a}.


\section{Adiabatic Quantum Computation}

In the previous sections it was shown for specific models that quantum phase transitions encountered along a straight-line path
can be avoided if a nonlinear path is chosen.
It would be great if this could be extended to adiabatic quantum computation, where for the nontrivial models numerical
calculations based on a straight-line interpolation point to an inverse scaling of the fundamental energy gap with the 
system size \cite{znidaric2005,znidaric2006a,latorre2004a,orus2004a,schuetzhold2006b}.
Not only would such a path usually admit a significantly shortened runtime of the adiabatic algorithm but also
concerning robustness against thermal excitations a constant lower bound on the fundamental energy gap would be 
highly desirable.
As a first step, we will show that there exist paths consisting of a series of straight-line interpolations, 
along which the fundamental gap can be calculated analytically.
As one paradigmatic example for the problem class NP, we will illustrate this for 
Exact Cover 3 (EC3), which is a special case of 3-SAT.
The EC3-problem can be defined as follows \cite{farhi2001}:
Given $m$ clauses each involving positions of three bits $C_i=(p_i^1, p_i^2, p_i^3)$ with $1\le i \le m$ and $1 \le p_i^\alpha \le n$ 
one is looking for the $n$-bit bitstring $\{b_1 \ldots b_n\}$ with $b_i \in \{0,1\}$ that fulfills for each clause
$b_{p_i^1} + b_{p_i^2} + b_{p_i^3}=1$, where '$+$' denotes the integer sum.
A problem Hamiltonian encoding the solution to the EC3 problem in its ground state can be given as \cite{banuls2006a,schuetzhold2006b}
\bea\label{Eexactcover}
H_{\rm P} &=& \sum_{i=1}^m \left[\f{1} - \hat{z}_{p_i^1} - \hat{z}_{p_i^2} - \hat{z}_{p_i^3}\right]^2
\eea
where $\hat{z}_k = \frac{1}{2}\left(\f{1}-\sigma^z_k\right)$.

Unfortunately, since the neighborship topology in the irregular lattice defined by $n_{ij}$ in an EC3 problem
cannot be represented by next neighbor interactions for a fixed dimensionality and does not even have a regular 
structure, there is no obvious unique way of defining a nonlinear interpolation scheme, compare also \cite{farhi0208135}.
Here, it will be proposed that by turning on the clauses one by one in (\ref{Eexactcover}) whilst maintaining
a unique ground state one should obtain a considerable improvement in the fundamental energy gap for the
average EC3 problem.
We define a partial problem Hamiltonian via
\bea\label{Epartial}
H_{\rm P}^k = \left[\f{1} - \hat{z}_{p_k^1} - \hat{z}_{p_k^2} - \hat{z}_{p_k^3}\right]^2
\eea
The series of Hamiltonians in the nonlinear path can then be written as
\bea\label{Ehamseries_ec3}
H_k = \f{1} - \ket{\Psi_k}\bra{\Psi_k}
\eea
connected by stepwise linear interpolations (\ref{Estepwise}) for 
\mbox{$0\le k \le m$} and where the
unique ground states are the symmetrized superpositions of solutions to a subset of clauses
\bea
\ket{\Psi_k} = \frac{1}{\sqrt{N_k}} \sum_{z \;:\; \left(H_{\rm P}^1 + \ldots + H_{\rm P}^k\right) \ket{z}=0} \ket{z}\,,
\eea
where we have denoted the ground state degeneracy of $H_{\rm P}^1 + \ldots + H_{\rm P}^k$ by $N_k$.
This of course does not uniquely define an interpolation scheme, since we may exchange the order of clauses without changing the final solution.
For the linear interpolation between two projector Hamiltonians
\mbox{$H_k(s) = \f{1} -(1-s) \ket{\Psi_k}\bra{\Psi_k} - s \ket{\Psi_{k+1}}\bra{\Psi_{k+1}}$}, the simple result for the adiabatic 
Grover algorithm \cite{roland2002} may be easily generalized \cite{wei2006a} to obtain the eigenvalues
\mbox{$\lambda_{0/1} = \frac{1}{2}\left(1\pm \sqrt{1-4s(1-s) \left(1-\abs{\braket{\Psi_k}{\Psi_{k+1}}}^2\right)}\right)$} and
\mbox{$\lambda_{2} = \ldots = \lambda_{2^n-1} = 1$},
such that the minimum fundamental energy gap equates to 
\mbox{$g_{\rm min}=\abs{\braket{\Psi_k}{\Psi_{k+1}}}=\sqrt{\frac{N_{k+1}}{N_{k}}}\equiv \sqrt{r_k}$}
(where $r_k \le 1$ is the reduction factor in the number of solutions at step $k$).
For an average EC3 problem with a unique solution, this would be a significant improvement, since 
typically a modest number of states will violate only a single clause.
For extreme problems however, an exponential number of basis states could possibly just violate only a single clause, such that even 
re-ordering of the clauses would not yield a polynomial scaling of the fundamental energy gap.
However, for direct interpolation between $H_0$ and $H_m$ (where we would reproduce the adiabatic Grover algorithm 
\cite{roland2002}) we obtain (for a problem with a unique solution such that $N_m=1$) 
the minimum fundamental gap of $g_{\rm min}^{\rm direct}=1/\sqrt{2^n}$, which is always significantly smaller than the smallest gap that could
be encountered with stepwise interpolation.

A further drawback of the scheme is that the Hamiltonians (\ref{Ehamseries_ec3}) are generally nonlocal and cannot be efficiently implemented, 
as can for example be seen from the formal representation of its ground state
\bea
\ket{\Psi_k} = \sqrt{\frac{N}{N_k}} \left[\prod_{\ell=1}^k \left(\f{1}-H_{\rm P}^\ell\right) \left(\f{4}-H_{\rm P}^\ell\right)\right] \ket{\cal S}\,.
\eea
Apart from the (generally unknown) normalization factor $N_k$, one would from inserting (\ref{Epartial}) in above equation obtain an 
exponential number of terms involving up to $n$-body interactions and also using three-body interactions in (\ref{Epartial}) as in the
original approaches \cite{farhi2001,orus2004a,latorre2004a} would not cure the problem.

However, it is also known that the conventional straight-line adiabatic algorithms with local Hamiltonians are more efficient than 
straight-line adiabatic algorithms that use a projection operator \cite{farhi_fail} -- even when both use the same initial and final ground states.
It would therefore be an interesting option to approximate the projection operators in (\ref{Ehamseries_ec3}) by few-body Hamiltonians 
with the same (or similar) ground states or to explore other nonlinear interpolation paths for local Hamiltonians.


\section{Summary}

For two analytically solvable models models with quantum phase transitions along a straight-line interpolation
we have shown that a nonlinear path provides a constant lower bound for the fundamental energy gap.
For closed quantum systems, this would enable a very efficient adiabatic preparation of highly entangled ground 
states \cite{dziarmaga2005a,cincio2007a}.
The used Hamiltonians in sections \ref{Sising} and \ref{Scluster} are always local and require local control of the couplings.
They do not require complicated time-dependencies 
of the coupling constants as in adiabatic rotation \cite{siu2007a} or with local adiabatic evolution \cite{roland2002,schaller2006a}.
For the quantum Ising model in a transverse field, the robustness of the scheme in the presence of a reservoir
is hampered by the two-fold degeneracy of the final ground state -- unless one is able to separate the even and odd 
subspaces by a safe energy barrier.
For the cluster model in a transverse field one does not have this problem, since the ground state is always unique.
Unfortunately, it requires 3-body interactions for the 1d lattice and 5-body interactions for the 2d lattice.
However, it is also known that the 2d cluster state can be approximated by the ground state of Hamiltonians with
two-body interactions only \cite{bartlett2006a}.
The approach could also be of interest for other models with just two-body interactions having highly entangled ground states
\cite{peschel2004a,camposvenuti2006a}.
Adiabatic quantum computation might also strongly benefit from using nonlinear paths in both algorithmic performance
and robustness against thermal excitations.
Beyond this, the presented idea may give rise to a class of Hamiltonians that can be used to calibrate numerical 
methods such as DMRG \cite{eisert2006b}.


\section{Acknowledgments}

The author gratefully acknowledges discussions with
R. Sch\"utzhold, T. Brandes, J. Eisert, I. Peschel, S. Mostame, and M. Cramer.

$^*$\,{\small\sf schaller@itp.physik.tu-berlin.de}



\begin{thebibliography}{9999}

\bibitem{shor1997}
P. W. Shor,
SIAM J. Comput. {\bf 26}, 1484-1509 (1997).

\bibitem{grover1997a}
L. K. Grover,
Phys. Rev. Lett. {\bf 79}, 325-328 (1997).

\bibitem{nielsen2000}
M.~A.~Nielsen and I.~L.~Chuang,
{\em Quantum Computation and Quantum Information},
Cambridge University Press, Cambridge (2000).

\bibitem{breuer2002}
H.-P. Breuer and F. Petruccione, 
{\em The Theory of Open Quantum Systems},
Oxford University Press, Oxford (2002).

\bibitem{raussendorf2001a}
R. Raussendorf and H. J. Briegel,
Phys. Rev. Lett. {\bf 86}, 5188 - 5191 (2001).

\bibitem{raussendorf2003a}
R. Raussendorf, D. E. Browne, and H. J. Briegel, 
Phys. Rev. A {\bf 68}, 022312 (2003).

\bibitem{pachos2001a}
J. Pachos and P. Zanardi, 
Int. J. Mod. Phys. B {\bf 15}, 1257 (2001).

\bibitem{farhi2001}
E.~Farhi {\em et al.},
Science {\bf 292}, 472 (2001);
E. Farhi {\em et al.},
{\tt arXiv:quant-ph/0001106v1} (2000).

\bibitem{sarandy2004}
M. S. Sarandy, L.-A. Wu, and D. A. Lidar,
Quant. Inf. Proc. {\bf 3}, 331-349 (2004).

\bibitem{aharonov2007}
D.~Aharonov {\em et al.},
SIAM J. Comput. {\bf 37}, 166-194 (2007).

\bibitem{childs2001}
A. M. Childs, E. Farhi, and J. Preskill,
Phys. Rev. A {\bf 65}, 012322 (2001).

\bibitem{schaller2008a}
G. Schaller and T. Brandes,
Phys. Rev. A {\bf 78}, 022106 (2008).

\bibitem{latorre2004a}
J. I. Latorre and R. Orus,
Phys. Rev. A {\bf 69}, 062302 (2004).

\bibitem{orus2004a}
R. Orus and J. I. Latorre,
Phys. Rev. A {\bf 69}, 052308 (2004).

\bibitem{sachdev2000}
S. Sachdev,
{\em Quantum Phase Transitions},
Cambridge University Press (2000).

\bibitem{znidaric2005}
M. Znidaric,
Phys. Rev. A {\bf 71}, 062305 (2005).

\bibitem{znidaric2006a}
M. Znidaric and M. Horvat,
Phys. Rev. A {\bf 73}, 022329 (2006).

\bibitem{zanardi2006a}
P. Zanardi and N. Paunkovic,
Phys. Rev. E {\bf 74}, 031123 (2006).

\bibitem{dziarmaga2005a}
J. Dziarmaga,
Phys. Rev. Lett. {\bf 95}, 245701 (2005).

\bibitem{cincio2007a}
L. Cincio {\em et al.}, 
Phys. Rev. A {\bf 75}, 052321 (2007).

\bibitem{mostame2007a}
S. Mostame, G. Schaller, and R. Sch\"utzhold
Phys. Rev. A {\bf 76}, 030304(R) (2007).

\bibitem{jordan1928a}
P. Jordan and E. Wigner,
Zeitschrift f\"ur Physik {\bf 47}, 631-651 (1928).

\bibitem{farhi0208135}
E. Farhi, J. Goldstone, and S. Gutmann,
{\tt arXiv:quant-ph/0208135v1} (2002).

\bibitem{siu2007a}
M. S. Siu,
Phys. Rev. A {\bf 75}, 062337 (2007).

\bibitem{porras2004a}
D. Porras and J. I. Cirac,
Phys. Rev. Lett. {\bf 92}, 207901 (2004).

\bibitem{mostame0803}
S. Mostame and R. Sch\"utzhold,
{\tt arXiv:0803.1093v2} (2008).

\bibitem{arpack1998}
R.~B.~Lehoucq {\em et al.} 
{\em ARPACK Users' Guide: Solution of Large-Scale Eigenvalue 
Problems with Implicitly Restarted Arnoldi Methods}, 
SIAM, Philadelphia, {\tt http://www.caam.rice.edu/software/ARPACK} (1998).

\bibitem{schaller2006a}
G. Schaller, S. Mostame, and Ralf Sch\"utzhold,
Phys. Rev. A {\bf 73}, 062307 (2006).

\bibitem{jansen2007a}
S. Jansen, M. B. Ruskai, and R. Seiler,
J. Math. Phys. {\bf 48}, 102111 (2007).

\bibitem{fubini2007a}
A. Fubini, G. Falci, and A. Osterloh,
N. J. Phys. {\bf 9}, 134 (2007).

\bibitem{balachandran2008a}
V. Balachandran and J. Gong,
Phys. Rev. A {\bf 77}, 012303 (2008).

\bibitem{doherty08024314}
A. C. Doherty and S. D. Bartlett,
{\tt arXiv:0802.4314v1} (2008).

\bibitem{pachos2004a}
J. K. Pachos and M. B. Plenio, 
Phys. Rev. Lett. {\bf 93}, 056402 (2004).

\bibitem{bartlett2006a}
S. D. Bartlett and T. Rudolph,
Phys. Rev. A {\bf 74}, 040302(R) (2006).

\bibitem{xu2004a}
C. Xu and J. E. Moore, 
Phys. Rev. Lett. {\bf 93}, 047003 (2004).

\bibitem{schuetzhold2006b}
R. Sch\"utzhold and G. Schaller,
Phys. Rev. A {\bf 74}, 060304(R) (2006).

\bibitem{banuls2006a}
M. C. Banuls {\em et al.}, 
Phys. Rev. A {\bf 73}, 022344 (2006).

\bibitem{roland2002}
J. Roland and N. J. Cerf,
Phys.\ Rev.\ {\bf 65}, 042308 (2002).

\bibitem{wei2006a}
Z. Wei and M. Ying,
Phys. Lett. A {\bf 356}, 312 (2006).

\bibitem{farhi_fail}
E. Farhi {\em et al.},
Int. J. Quant. Inf. {\bf 6}, 503-516 (2008).

\bibitem{peschel2004a}
I. Peschel,
J. Stat. Mech.: Theor. Exp., P12005 (2004).

\bibitem{camposvenuti2006a}
L. Campos Venuti, C. Degli Esposti Boschi, and M. Roncaglia,
Phys. Rev. Lett. {\bf 96}, 247206 (2006).

\bibitem{eisert2006b}
J. Eisert,
Phys. Rev. Lett. {\bf 97}, 260501 (2006).

\end{thebibliography}
\end{document}